\documentclass[9pt,twocolumn,twoside]{opticajnl}
\journal{opticajournal} 

\setboolean{shortarticle}{true}

\usepackage{lineno}
\usepackage{comment}
\usepackage{amsfonts}
\usepackage{amssymb}
\usepackage{mathtools}
\usepackage{braket}
\usepackage{float}
\usepackage{amsmath}
\newcommand\omicron{o}
\title{Power output optimization in complex laser systems by means of polarization control}

\author[1,2,*]{Dominika~Jochcov\'{a}}
\author[1]{Ond\v{r}ej~Slez\'{a}k}
\author[2]{Ivan~Richter}
\author[1]{Martin~Smr\v{z}}
\author[1]{Tom\'{a}\v{s}~Mocek}

\affil[1]{HiLASE Centre, Institute of Physics of the Czech Academy of Sciences, Za Radnic\'{i} 828, 252 41, Doln\'{i} B\v{r}e\v{z}any, Czech Republic}
\affil[2]{Faculty of Nuclear Sciences and Physical Engineering, Czech Technical University in Prague, B\v{r}ehov\'{a} 7, 115 19, Prague, Czech Republic}

\affil[*]{dominika.jochcova@hilase.cz}

\begin{abstract}
Recently, the polarimetric method for thermally-induced polarization changes driven power losses (TIPCL) mitigation in complex laser systems has been developed. However, the final optimization relied on the four-parameter numerical process. This letter provides a fully analytical direct calculation alternative to this optimization process. The validity of this approach is demonstrated on the previously published data from pulsed laser system Bivoj/DiPOLE100. The new approach provides a deeper insight into the polarimetric method for TIPCL suppression and also brings a more precise, reliable, and faster alternative to the numerical process used earlier.  
\end{abstract}
\setboolean{displaycopyright}{false} 
\begin{document}
\maketitle


A recent development on the field of thermal-stress-induced birefringence (TSIB) compensation in complex laser systems assisted by polarimetric measurement \cite{Slezak2022} led to the formulation of the optimization problem essential for a successful usage of such method. This optimization step can be formulated as follows: \emph{one is searching for the uniform polarization state suffering the lowest distortion when it propagates through defined non-uniformly polarizing optical system in the sense that its part blocked by the output elliptical polarizer carries minimal energy. The procedure should also determine the parameters of such output elliptical polarizer.} The optimisation process is therefore searching for four parameters of the system. The azimuth and the phase delay of the input polarization and the same couple of parameters of the output polarizer. So far, the problem has been successfully solved numerically in \cite{Slezak2022}. The numerical solution is, however, relatively slow and does not provide any more general insight into the problem. For example, the ideal configuration for a simple systems for which an analytical solution TSIB is available, e.g. heated cylindrically symmetrical rod amplifier, cannot be deduced on the basis of the numerical solution. Also the limits of usability of such optimization are not quite clear from the numerical solution. All these uncertainties led to the searching for an analytical solution presented in this letter.

\begin{figure}[ht]
\centering
\includegraphics[width=.8\linewidth]{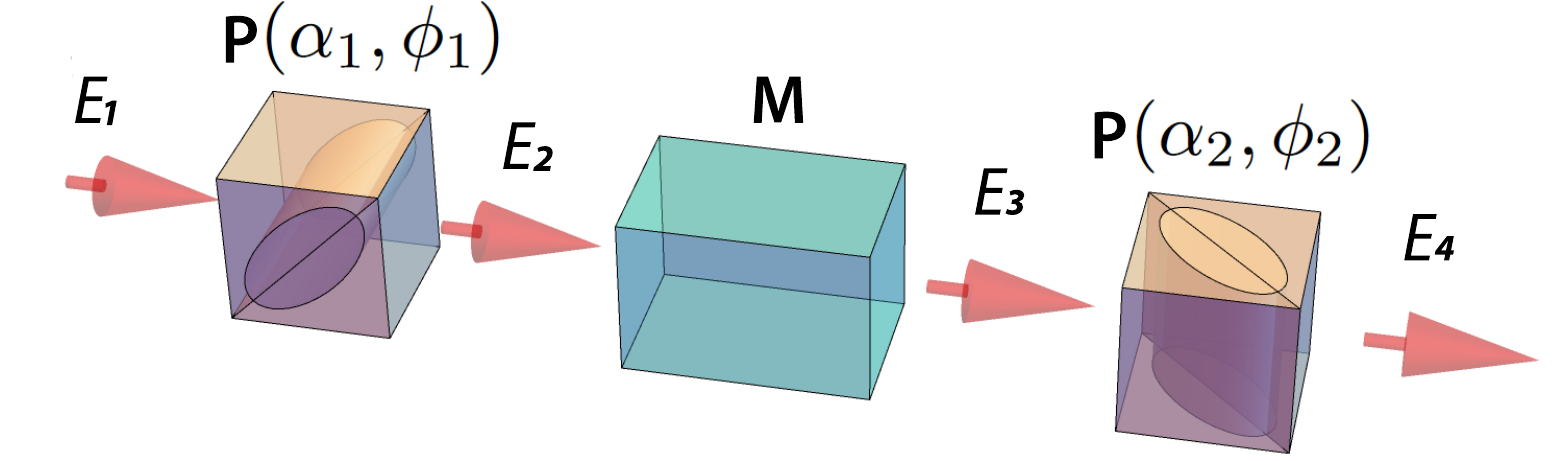}
\caption{The schematic of generalized polarizer-analyzer setup for TIPCL evaluation; distorting optical system $\mathbf{M}$ is placed between two elliptical polarizers $\mathbf{P}$.}
\label{fig:scheme}
\end{figure}

The volumetric heat load generated in the laser amplifiers and other optical components give rise to the distortions of the passing-through laser beam. Among them, TSIB induces non-uniform polarization changes across the beams' cross-section. Such effect can significantly decrease the efficiency of the polarization-sensitive phenomena, e.g. higher harmonics conversion \cite{Bayramian2016,Phillips2021,Divoky2023,Clarke2023}, optical parametric amplification \cite{Bromage2019} or make impossible polarization-sensitive experiments. In combination with a linear diattenuator, typically polarizer, a significant loss of energy also occurs accompanied by the intensity profile change. Such loss of energy can theoretically reach up to 50\%. However, the typical values reported from the high-energy high-average-power laser systems lie, from our experience, usually between 20\% and 40\%. As reported in \cite{Slezak2022}, the Mueller-matrix polarimetry based method of TIPCL reduction led to the decrease of the power loss from 33\% to 9\%, and after the more precise measurement even down to 3\% \cite{Divoky2023}. It was also shown that the optical setup used for the compensation is equivalent to the arrangement in which the whole amplifier chain is placed between two properly chosen elliptical polarizers. Such equivalent scheme is shown in Fig.~\ref{fig:scheme}. 
The azimuth angles $\alpha_1$, $\alpha_2$ and the phase delays $\phi_1$, $\phi_2$ are then subject to the above mentioned optimization process which is within this letter transformed into the direct calculation of these four parameters. Such an approach makes the optimization much faster, precise, and also provides the novel insight into the process of the compensation itself. 

The analysis of the problem formulated at the beginning of this section is addressed in 3 gradual steps. The first step specifies the parameters of the output elliptical polarizer for a given general input polarization state. The parameters of the output polarizar are, in this case, expressed as functions of the input polarizer characteristics. The second step provides the values of the input polarizer parameters suffering the minimal polarization distortion for a given optical system. The third step summarizes the formulation of the method in the Mueller calculus, thus allowing direct connection of the developed optimization method with an experimental Mueller matrix of the studied optical system obtained from the polarimetric measurement.

The first step of the process is dedicated to the specification of the output elliptical polarizer for the fixed input polarization state. Let us use the standard parametrization of the Jones vector representing the totally-polarized input state of light as $ E(\alpha,\phi)=\left[E_x,E_y\right]^T=\left[\cos{\alpha}e^{i\phi},\sin{\alpha}\right]^T$,
where $\alpha \in (0,\frac{\pi}{2})$ is the azimuth of the polarization ellipse and $\phi \in (-\pi,\pi)$ is the phase delay between orthogonal-base linear polarizations. Let us assume that the considered optical system is generally birefringent i.e. represented as a non-depolarizing optical system, possessing no diattenuation. The equivalence theorems of the Jones calculus proved in \cite{HurwitzJr.1941,Whitney1971} allow us to represent the above-defined optical system by a single unitary Jones matrix $\mathbf{M}(x,y)$ with transversely (with respect to the beam propagation direction) dependant elements $m_i\left(x,y\right) $ and $
 \varphi\left(x,y\right)$ parameterized as
\begin{equation}
    \mathbf{M}=e^{i\varphi/2}\begin{bmatrix}
        m_1 & m_2 \\
        -m_2^* & m_1^* \\
    \end{bmatrix},
    \label{eq_munit}
\end{equation}
where $|m_1|^2+|m_2|^2=1$. It should be noted that isotropic absorption or amplification does not influence the polarization state and therefore it does not have to be taken into consideration. The output polarization state, leaving the optical system $\mathbf{M}\left(x,y\right)$, is described by a transversely non-uniform Jones vector $E_3=E_3\left(\alpha\left(x,y\right),\phi\left(x,y\right)\right)$. The output polarizer $\mathbf{P}\left(\alpha_2,\phi_2\right)$ (see Fig.~\ref{fig:scheme}) is represented by an orthogonal projector to the state $E\left(\alpha_2,\phi_2\right)$, represented by a Hermitian idempotent matrix
\begin{equation}
\mathbf{P}\left(\alpha_2,\phi_2\right)=E\left(\alpha_2,\phi_2\right) E^{\dagger}\left(\alpha_2,\phi_2\right), 
    \label{eq_p}
\end{equation}
where $E^{\dagger}$ is the conjugate transpose of the vector $E$. For practical reasons, let us divide the output transverse plane into a set of discrete $N \times M$ points and consider a constant Jones vector in each point of these node points. This leads to the discretized quantities $E_3(\alpha (x,y),\phi(x,y)) \rightarrow E_3(\alpha (x_i,y_j),\phi(x_i,y_j)) \equiv E_3(x_i,y_j)$, where $i \in \{1,2,\dotsb, N\}$, $j \in \{1,2,\dotsb, M\}$. This discretization is advantageous for the work with the experimental data, which are inherently represented by a discrete set of values. Also the numerical approaches, which are mostly used for the evaluation of the induced birefringence effects, are typically performed on a discrete grid. Each point $(x_i,y_j)$ represents an area upon which the Jones vector is considered to be constant. When light passes the output polarizer, the intensity of the passed-through beam is determined by $I\left(x_i,y_j\right)=\left|\mathbf{P}E_3\left(x_i,y_j\right)\right|^2$. For the further development of the method, it is beneficial to define the sum output intensity of the beam $I_{tot}$ over all node points as
\begin{equation}
    I_{tot}=\sum_{i=1}^{N}\sum_{j=1}^{M}\left|\mathbf{P}E_3\left(x_i,y_j\right)\right|^2.
    \label{eq_tot}
\end{equation}
The objective is to maximize the sum intensity $I_{tot}$ by the selection of an appropriate output polarizer $\mathbf{P}$. It can be derived (see Appendix~A) that there exists an equivalent expression of Eq.  (\ref{eq_tot}) in the form
\begin{equation}
I_{tot}=P^\dagger \mathbf{D} P,
    \label{eq_I}
\end{equation}
where $P$ stands for the eigenstate of the polarizer $\mathbf{P}$ and $\mathbf{D}$ is the matrix defined as
\begin{equation}
\mathbf{D}\equiv \sum_{i=1}^{N}\sum_{j=1}^{M} E_3(x_i,y_j)E_3(x_i,y_j)^\dagger .
    \label{eq_d}
\end{equation}
$\mathbf{D}$ is a finite sum of Hermitian idempotent matrices that represent projectors onto output polarization states. Consequently, $\mathbf{D}$ is Hermitian, although it is generally not idempotent, thus representing a general diattenuator (also called partial polarizer). Let $d_1$ and $d_2$ be the eigenvalues of $\mathbf{D}$. The output sum intensity obviously varies with the choice of the output polarizer $\mathbf{P}$. However, the eigenvalues $d_1$ and $d_2$ serve as limits on $I_{tot}$ ensuring that $I_{tot}$ consistently lies within the interval $I_{tot}\in \left(d_2,d_1\right)$ where, without loss of generality, $d_{1,2}$ fulfill $d_1 \geq d_2$. Let us denote the eigenstates of $\mathbf{D}$ as $D_{1,2}$, i.e. $\mathbf{D}D_{1,2}=d_{1,2}D_{1,2}$. Apparently, maximization of the output power of the laser beam can be achieved by the proper choice of the eigenstate of the output polarizer $\mathbf{P}$ as $P\equiv D_1$.

Minimal dimensionless depolarization loss, according to its common laser physics definition as a ratio of power leakage through the output polarizer to the total power, $\eta_{min}$ is related to the eigenvalues of $\mathbf{D}$ as
\begin{equation}
    \eta_{min}=\frac{d_2}{M N}=\bigg(1-\frac{d_1}{M N}\bigg) .
\end{equation}

The second step of the process deals with searching for the optimal parameters of the input elliptical polarizer $\mathbf{P}\left(\alpha_1,\phi_1\right)$ which produces the input polarization state for the studied optical system. A widely used method for representing the polarization of quasi-monochromatic waves is based on the coherency matrix \cite{Wolf2008}. However, the coherency matrix representing fully polarized, partially correlated fields with a spatially varying polarization state can also be constructed \cite{Tervo2009} in a similar way. In terms of the spatially varying elements of the Jones vector, the polarization matrix $\mathbf{J}$ is defined as
\begin{equation}
\mathbf{J}= \begin{bmatrix}
    \langle E_xE_x^* \rangle & \langle E_xE_y^* \rangle  \\
    \langle E_yE_x^*  \rangle & \langle E_yE_y^* \rangle 
\end{bmatrix} = \begin{bmatrix}
    j_{xx} & j_{xy} \\
    j_{yx} & j_{yy}
\end{bmatrix},
\label{eq_jmat}
\end{equation}
where the angular brackets represent the spatial mean value over the cross-section of the beam. Thus, considering $E \equiv E_{3}$, the $\mathbf{J}$-matrix characterizes the statistical properties of the spatially resolved polarization state, i.e. the non-uniformly totally polarized (NUTP) beam emerging from the optical medium \cite{Piquero2020}. Let $E_2$ be the polarization state entering the optical system, thus $E_2=E_2\left(\alpha_1,\phi_1\right)$ is the eigenstate of input polarizer $\mathbf{P}\left(\alpha_1,\phi_1\right)$. The NUTP beam $E_3$ leaving the optical system is given as $E_3\left(x,y\right)=\mathbf{M}\left(x,y\right)E_2\left(\alpha_1,\phi_1\right)$. By the substitution for $\mathbf{M}$ from (\ref{eq_munit}) and for the eigenvector of $\mathbf{P}\left(\alpha_1,\phi_1\right)$, one gets
\begin{equation}
   E_3=e^{i\varphi/2}\begin{bmatrix}
        m_1 \cos \alpha_1 e^{i\phi_1}+m_2\sin \alpha_1 \\
        -m_2^* \cos \alpha_1 e^{i\phi_1}+m_1^* \sin \alpha_1
    \end{bmatrix}.
    \label{eq_eout}
\end{equation}
Taking into account that the NUTP beam represents the mixture of many polarization states over the cross-section of the beam, the spatial degree of polarization $p$ can be defined for a NUTP beam. It should be noted here that from the point of view of the studied problem, the beams with an arbitrary spatial distribution of the polarization states over the beam are indistinguishable as long as the total area occupied by every single polarization state remains the same. In other words, every two NUTP beams with the same composition of the polarization states regardless of their spatial distribution will lead to exactly the same resulting power loss and therefore they can be exchanged freely. One of these equivalent NUTP beams is always the one with a random spatial distribution of the desired polarization states composition. Such a beam is the one for which it makes sense to define the degree of polarization analogously to the partially polarized light. The degree of polarization $p$ can be then expressed directly from the polarization matrix (\ref{eq_jmat}) as \cite{Mandel2013}
\begin{equation}
    p^2=\text{Tr}^2\mathbf{J}-4\text{Det}\mathbf{J} .
    \label{eq_dettr}
\end{equation}
Since $\mathbf{M}$ was assumed to be the unitary Jones matrix, the trace of $\mathbf{J}$-matrix is identically equal to unity $\text{Tr}\mathbf{J}=\langle |m_1|^2 + |m_2|^2 \rangle=1$.

The figure of merit of the studied problem is the quantity usually called the depolarization loss $\eta$. When obtained from the experimental data, the depolarization loss is calculated as the ratio of the power leakage through the optical system placed between crossed polarizers to the total power transmitted through the optical system without these polarizers. The depolarization loss $\eta$ is directly connected to the spatial degree of polarization. Let us consider $I_p$ and $I_{np}$ be the completely polarized and completely unpolarized intensity compounds of the beam, respectively, i.e. $I=I_p+I_{np}$. When a beam is transmitted through the output polarizer $\mathbf{P}\left(\alpha_2,\phi_2\right)$, the transmittance $T(I_p)=1$ for a proper choice of the eigenstate $\left(\alpha_2,\phi_2\right)$. In contrary, $T(I_{np})=1/2$ for an arbitrary choice of $\mathbf{P}\left(\alpha_2,\phi_2\right)$. Consequently, maximal transmittance is given by
\begin{equation}
    T(I_p+I_{np})=1-\frac{I_{np}}{2(I_{np}+I_p)}.
\end{equation}
Since $p=I_p/(I_p+I_{np})$, the depolarization loss can be expressed as
\begin{equation}
    \eta=(1-T) = \frac{1}{2}(1-p).
\end{equation}
The depolarization loss $\eta$ reaches its minimum as $p$ approaches its maximal value. Substituting (\ref{eq_eout}) into (\ref{eq_jmat}) and (\ref{eq_dettr}) and simplifying the expression, one obtains
\begin{align}
    \nonumber p^2 &=1-\frac{1}{2}(1-\cos 4\alpha_1)(\mu-2\cos 2\phi_1 \mathfrak{Re}\{\nu\}+\\
    \nonumber&+ 2\sin 2\phi_1 \mathfrak{Im}\{\nu\})-2\sin 4\alpha_1(\cos \phi_1 \mathfrak{Re}\{\xi\}+\\
    &+\sin \phi_1 \mathfrak{Im}\xi \})-\omicron(1+\cos 4 \alpha_1)-\lambda,
    \label{eq_4det}
\end{align}
where symbols $\mathfrak{Re}\{ \}$, $\mathfrak{Im}\{ \}$ stand for real resp. imaginary part, optical system parameters $\lambda,\mu,\nu,\xi,\omicron$ are given as
\begin{align}
    \nonumber\lambda &=2\langle |m_1|^2\rangle \langle |m_2|^2 \rangle ,\\
    \nonumber\mu &=\langle |m_1|^2\rangle ^2+\langle |m_2|^2\rangle ^2-2|\langle m_1 m_2^*\rangle|^2-|\langle m_1^2\rangle|^2-|\langle m_2^2\rangle|^2 ,\\
    \nonumber\nu &=\langle m_1 m_2^*\rangle ^2-\langle m_1^2 \rangle \langle m_2^{*2} \rangle ,\\
    \nonumber\xi &=\langle m_1^{*2}\rangle \langle m_1 m_2 \rangle -\langle m_1^{*} m_2^{*}\rangle \langle m_2^2 \rangle + \langle |m_2|^2-|m_1|^2\rangle \langle m_2 m_1^* \rangle ,\\
    \omicron &=\langle |m_1|^2\rangle \langle |m_2|^2 \rangle-2|\langle m_1 m_2\rangle|^2.
    \label{eq_parameters}
    \end{align}
The squared spatial degree of polarization $p^2$ (\ref{eq_4det}) is a function of two input polarization state variables $\left(\alpha_1,\phi_1\right)$ and parameters of the optical system (\ref{eq_parameters}) which are fixed by the studied optical system. Let us examine the stationary points of $p^2=p^2\left(\alpha_1,\phi_1\right)$ by the search for the zeros of its gradient $\nabla p^2=\mathbf{0}$. After the substitution $x \equiv \tan \phi_1$, the gradient equation is transformed into a 5th order polynomial in $x$ while $\alpha_1$ can be excluded from the equations. Two roots of this polynomial are always equal $\pm i$. This fact allows the decrease of the order of the polynomial to 3rd $\sum_{i=0} ^{3} a_ix^i =0$, where $a_i$ are the parameters depending exclusively on the optical system parameters (full expressions for $a_i$ are provided in a supplementary file). Taking into account the periodicity of $p^2\left(\alpha_1,\phi_1\right)$, its extremal values should be found in the interval $(0,\frac{\pi}{2})\times (-\pi,\pi)$. These values then define the optimal polarization state which suffers the lowest possible depolarization loss.

 In the third step of the process, the part of the calculation is transferred to the Mueller matrix formalism which is more preferable for the processing of the experimental data. The above derived technique will be transformed from the Jones notation to Mueller matrices. Such step makes this technique more applicable to the experimentally obtained results. Let us define the $W$ vector, which characterizes the general fully polarized state, as a Kronecker product of two Jones vectors in the form used in the previous sections
\begin{equation}
    W=E^*\otimes E.
    \label{eq_wvec2}
\end{equation}
The similar transformation can be done with the Jones matrix $\mathbf{M}$ given by (\ref{eq_munit}). Since the elements of $\mathbf{M}$ are already the functions of the spatial coordinates, one should, similarly to (\ref{eq_parameters}), apply the spatial mean value represented by the angular brackets. Let us denote the resulting matrix as $\mathcal{C}$, 
\begin{equation}
\mathcal{C}=\langle \mathbf{M}^*\otimes \mathbf{M} \rangle=    
    \begin{bmatrix}
      \langle |m_1|^2 \rangle  &{\langle m_1^*m_2 \rangle}  & c_{12}^* & \langle |m_2|^2 \rangle \\
      {-\langle m_1^*m_2^* \rangle}  & {\langle m_1^{*2} \rangle} & {-\langle m_2^{*2} \rangle} & {-c_{21}} \\
      {c_{21}^*}  & {c_{23}^*} &{c_{22}^*} & {-c_{21}^* }\\
      c_{14} & -c_{12} & -c_{12}^* & c_{11} \\
    \end{bmatrix}.
    \label{eq_covariance1}
\end{equation}
The connection to the experimentally preferred Mueller matrix notation is now becoming more obvious while there is a simple connection between the $W$ vector and the corresponding Stokes vector $S$ and also between the $\mathcal{C}$ matrix and the Mueller matrix $\mathcal{M}$.  
\begin{equation}
 S=\mathcal{A}W^*, \quad \mathcal{M}=\mathcal{A} (\mathbf{M}\otimes \mathbf{M}^*) \mathcal{A}^{-1}, 
 \label{eq_ms}
\end{equation}
where $\mathcal{A}$ matrix is defined as $\mathcal{A}=[\sigma_0,\sigma_1,\sigma_2,\sigma_3]^\dagger$, where $\sigma_i$ are the Pauli matrices written as a column vector \cite{Sheppard2016}. The spatial mean value can be then easily applied to the Mueller matrix $\left\langle\mathcal{M}\right\rangle=\mathcal{A}\mathcal{C}^*\mathcal{A}^{-1}$. The squared degree of polarization (\ref{eq_dettr}) can be, in this case, calculated as  
\begin{align}
    \nonumber p^2&=(q_1+q_4)^2-4(q_1q_4-q_2q_3)= \\
   & =(q_1-q_4)^2+4q_2q_3,
   \label{eq_pmue}
\end{align}
where $Q\equiv\mathcal{C}W=[q_1,q_2,q_3,q_4]^\intercal$ is a function of the input polarization parameters $\left(\alpha_1,\phi_1\right)$ and $\mathcal{C}$ is calculated from the experimentally obtained Mueller matrix $\mathcal{M}$. This formulation offers a straightforward means to compute the degree of polarization by direct utilizing the results from the experimental characterization of the optical system in the form of the Mueller matrix $\mathcal{M}$. 

\begin{figure}[h!tb]
    \centering    
    \includegraphics[width=.8\linewidth]{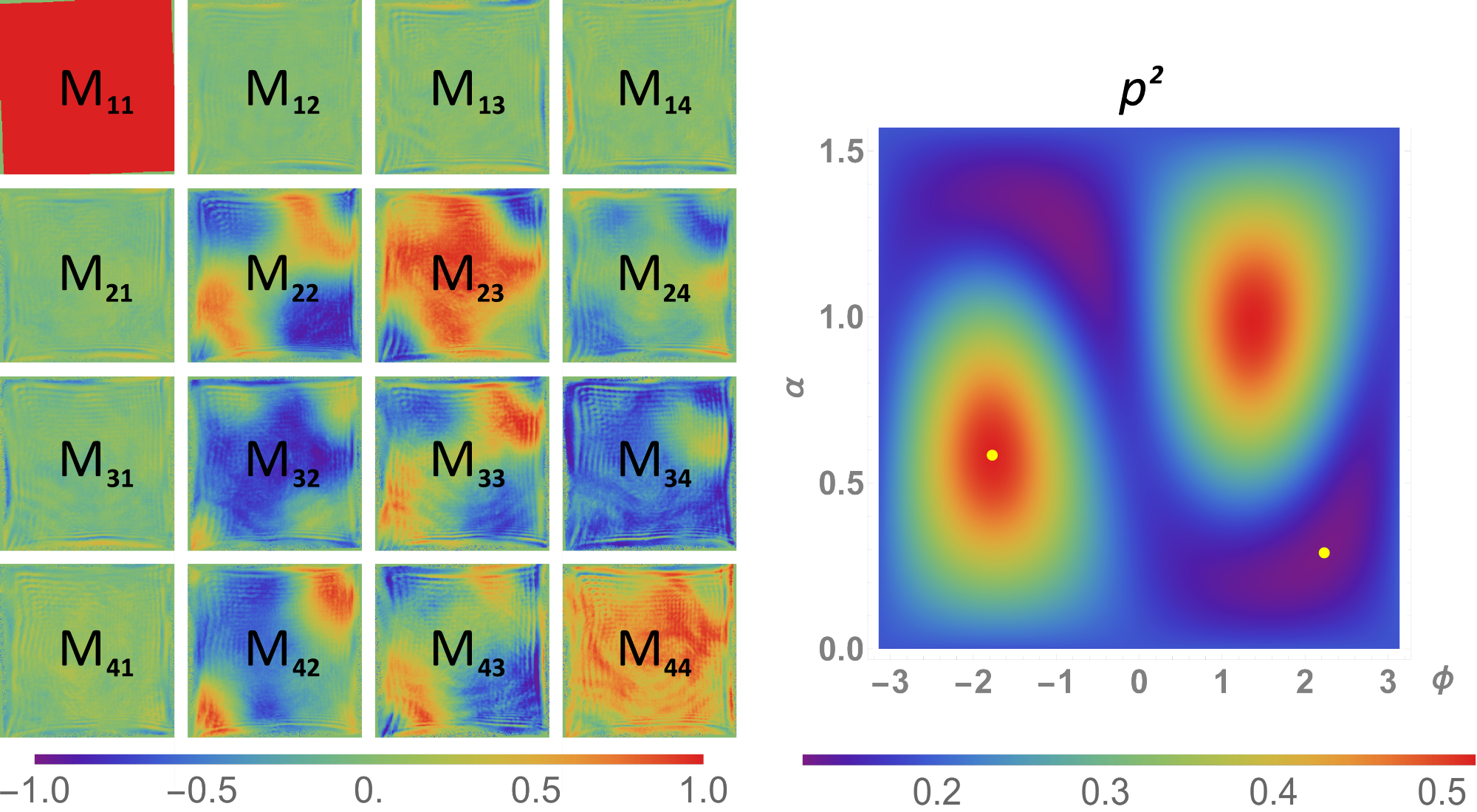}
    \caption{Spatially-resolved Mueller matrix of the laser system obtained from polarimetric measurement (left) and visualisation of the $p^2(\alpha_1,\phi_1)$ function calculated for the system (right).}
    \label{fig:ef}
\end{figure}

The above derived procedure will be demonstrated on the results presented in \cite{Slezak2022}. The subject to the spatially resolved Mueller matrix measurement was a last stage of the power amplifiers of the 100J/10Hz pulsed laser system "Bivoj/DiPOLE100" operated in HiLASE Centre, Czech Republic. The resulting Mueller matrix, presented also in \cite{Slezak2022}, is shown in Fig.~\ref{fig:ef}.
In the original reference, the numerical four parameter optimization has been used to find the ideal adjustment of the waveplates surrounding the amplifier chain. Let us demonstrate the direct calculation method on this example.

\begin{figure}[h!tb]
    \centering
    \includegraphics[width=.8\linewidth]{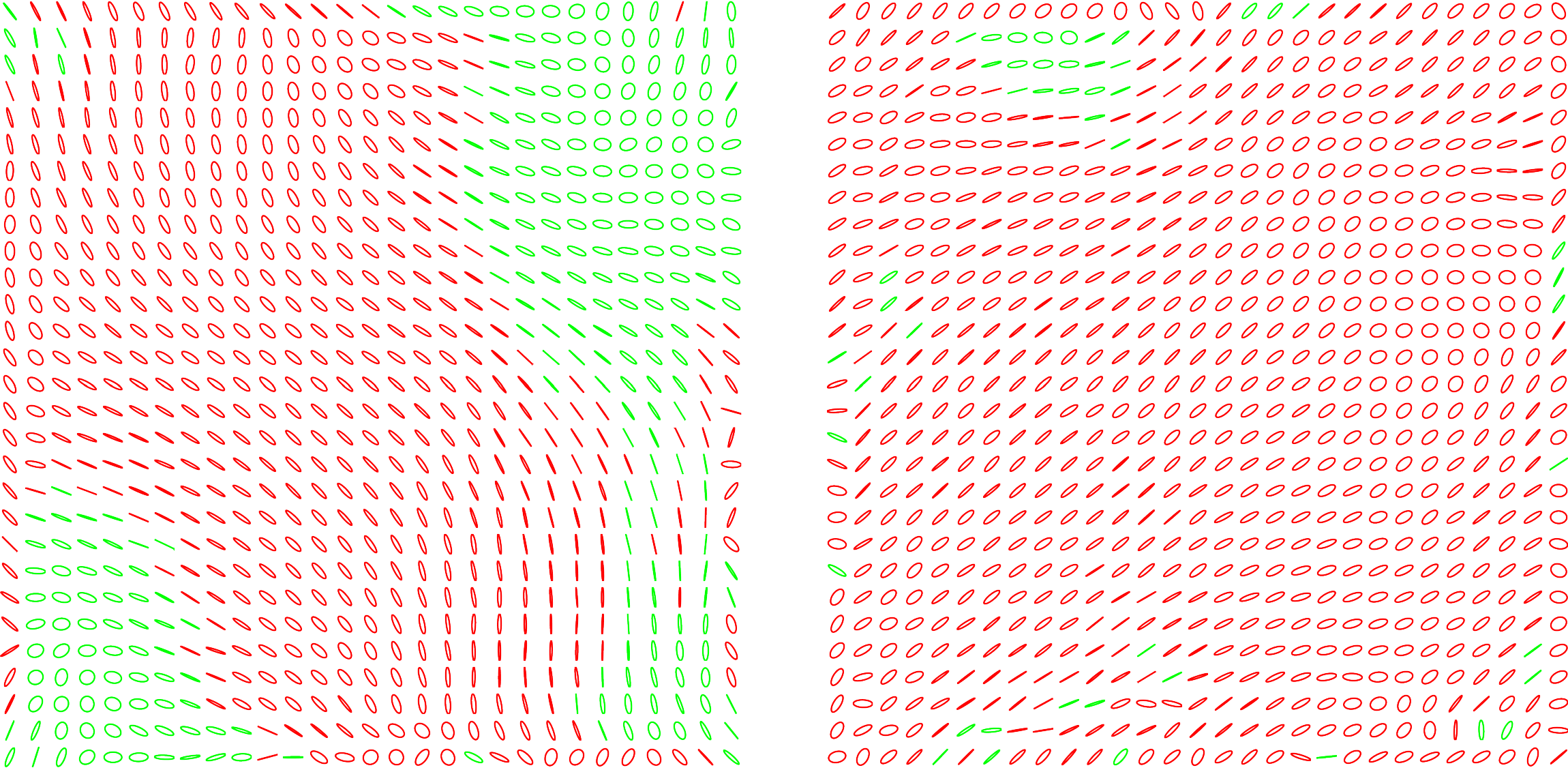}
    \caption{Polarization pattern of the cross-section of the output NUTP beam; for input linear horizontal polarization (left) and for optimal input polarization (right). The sense of rotation of the polarization ellipse is color-separated by green (clockwise) and red (anti-clockwise).}
    \label{fig:elip}
\end{figure}

\begin{figure}[h!tb]
    \centering
    \includegraphics[width=.8\linewidth]{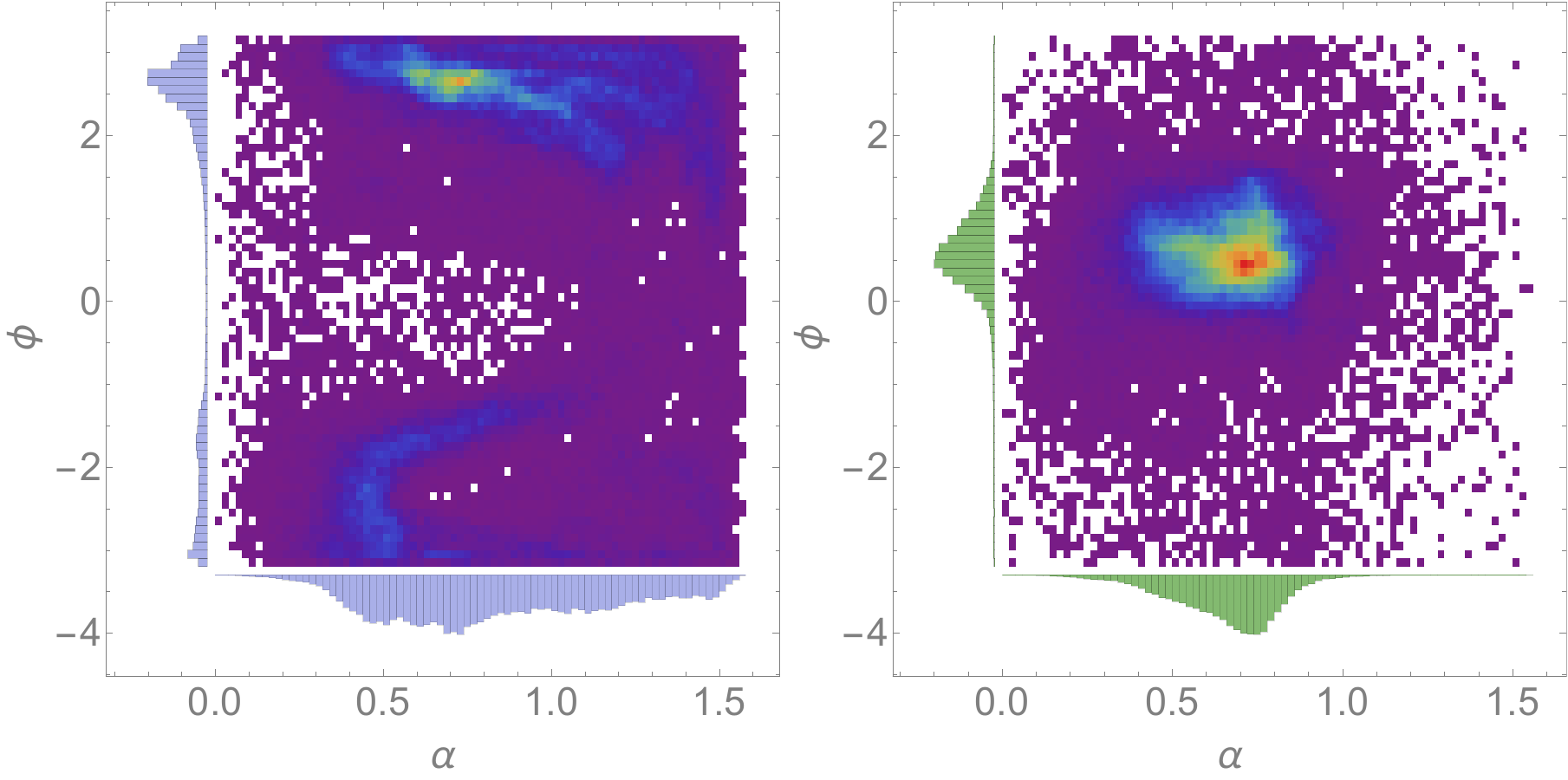}
    \caption{The density of states for output polarization vector $E_3$ in polarimetric coordinates $(\alpha,\phi)$. Input linear horizontal polarization (left) and optimized input polarization state (right).}
    \label{fig:hist}
\end{figure}

The optimal input polarization state can be directly computed from the Mueller matrix $\mathcal{M}(x,y)$ (Fig.~\ref{fig:ef}) of the amplifier chain by finding the maximum of $p^2$ (\ref{eq_pmue}). The $p^2$ as a function of the input polarization polarimetric coordinates $\left(\alpha,\phi\right)$ is shown in Fig.~\ref{fig:ef}. The polarization profile of the NUTP beam leaving the amplifier chain is then shown in Fig.~\ref{fig:elip}. 
The figure on the left demonstrates the NUTP beam leaving the amplifier with the linear horizontal polarization on the input while the right hand side profile is a NUTP beam produced by the amplifier from the ideal input polarization which can be read out from $p^2$ function as $\left(\alpha_1,\phi_1\right)=\left(0.584,-1.772\right)$ rad.
The degree of purity of the polarization state can be also well visualized by the density of polarization states in the polarimetric coordinates $\left(\alpha,\phi\right)$ as shown in Fig.~\ref{fig:hist}. 
The polarization states from Fig.~\ref{fig:elip} are shown. The transformation from the spatial to the polarimetric coordinates has been applied. The better localization of the polarization states in the second beam allows a more effective transmission by means of a properly chosen elliptical polarizer. The knowledge of the output NUTP beam allow us to construct the $\mathbf{D}$ operator according to (\ref{eq_d}). The output polarizer providing maximum transmission is the one whose eigenstate is exactly the same as the eigenstate of the $\mathbf{D}$ operator corresponding to the larger eigennumber $d_1$. The polarimetric coordinates of this eigenstate results to $\left(\alpha_2,\phi_2\right)=\left(0.693,0.997\right)$ rad. Together with the coordiantes of the input state these values are in excellent agreement with the ones obtained in \cite{Slezak2022}.

\subsection{Appendix}
The most suitable way how to derive the transition from the output polarizer applied to the NUTP beam produced by some optical system (\ref{eq_I}) is the usage of Dirac notation applied to the Jones vectors and matrices. Let
\begin{equation*}
    E_3(x_i,y_j)\rightarrow \ket{E_{ij}}, \quad P\rightarrow \ket{P},  \quad\mathbf{P}\rightarrow \ket{P}\bra{P},
\end{equation*}
consequently, (\ref{eq_tot}) takes the form
\begin{equation}
   \sum_{i=1}^{N} \sum_{j=1}^{M} |\mathbf{P}\ket{E_{ij}}|^2=\sum_{i=1}^{N}\sum_{j=1}^{M}|\ket{P}\braket{P|E_{ij}}|^2 ,
    \label{eq_g}   
\end{equation}
expanding the formula and considering the normalization $\braket{P|P}=1$:
\begin{equation}
\sum_{i=1}^{N}\sum_{j=1}^{M}\braket{E_{ij}|P}\braket{P|P}\braket{P|E_{ij}}=\sum_{i=1}^{N}\sum_{j=1}^{M}\braket{P|E_{ij}}\braket{E_{ij}|P}   ,
\end{equation}
the linearity of the scalar product allows for the change in the order of summation and multiplication
\begin{equation}
\bra{P}\bigg(\sum_{i=1}^{N}\sum_{j=1}^{M}\ket{E_{ij}}\bra{E_{ij}}\bigg)\ket{P}=\bra{P}\mathbf{D}\ket{P}.   
\label{eq_Ddef}
\end{equation}
Formula (\ref{eq_Ddef}) directly defines the formal partial polarizer $\mathbf{D}$.

\begin{backmatter}
\bmsection{Funding} Horizon 2020 Framework Programme (739573); European Regional Development Fund and the state budget of the Czech Republic project HiLASE CoE (CZ.02.1.01/0.0/0.0/15\_006/0000674); LasApp CZ.02.01.01/00/22\_008/0004573. Student Grant Competition of CTU in Prague (No. SGS22/185/OHK4/3T/14).


\bmsection{Disclosures} The authors declare no conflicts of interest.

\bmsection{Data availability} No data were generated or analyzed in the presented research.
\bmsection{Supplemental document} See Supplement 1 for supporting content.
\end{backmatter}

\bibliography{reference}

\begin{thebibliography}{10}
\newcommand{\enquote}[1]{``#1''}

\bibitem{Slezak2022}
O.~Slez{\'{a}}k, M.~Sawicka-Chyla, M.~Divok{\'{y}}, \emph{et~al.}, \enquote{{Thermal-stress-induced birefringence management of complex laser systems by means of polarimetry},} {\protect\JournalTitle{Scientific Reports}} \textbf{12}, 18334 (2022).

\bibitem{Bayramian2016}
A.~Bayramian, R.~Bopp, B.~Deri, \emph{et~al.}, \enquote{High-energy diode-pumped solid-state laser (dpssl) for high-repetition-rate petawatt laser systems,} in \emph{High-Brightness Sources and Light-Driven Interactions,}  (Optica Publishing Group, 2016), p. HT1B.5.

\bibitem{Phillips2021}
J.~P. Phillips, S.~Banerjee, P.~Mason, \emph{et~al.}, \enquote{{Second and third harmonic conversion of a kilowatt average power, 100-J-level diode pumped Yb:YAG laser in large aperture LBO},} {\protect\JournalTitle{Optics Letters}} \textbf{46}, 1808 (2021).

\bibitem{Divoky2023}
M.~Divoky, J.~Phillips, J.~Pilar, \emph{et~al.}, \enquote{{Kilowatt-class high-energy frequency conversion to 95 J at 10 Hz at 515 nm},} {\protect\JournalTitle{High Power Laser Science and Engineering}} \textbf{11}, 4--7 (2023).

\bibitem{Clarke2023}
D.~Clarke, J.~Phillips, M.~Divoky, \emph{et~al.}, \enquote{{Improved stability second harmonic conversion of a diode-pumped Yb:YAG laser at the 0.5 kW level},} {\protect\JournalTitle{Optics Letters}} \textbf{48}, 6320 (2023).

\bibitem{Bromage2019}
J.~Bromage, S.~W. Bahk, I.~A. Begishev, \emph{et~al.}, \enquote{{Technology development for ultraintense all-OPCPA systems},} {\protect\JournalTitle{High Power Laser Science and Engineering}} \textbf{7}, 1--11 (2019).

\bibitem{HurwitzJr.1941}
H.~{Hurwitz Jr.} and R.~C. Jones, \enquote{{A New Calculus for the Treatment of Optical Systems II. Proof of Three General Equivalence Theorems},} {\protect\JournalTitle{Journal of the Optical Society of America}} \textbf{31}, 493--499 (1941).

\bibitem{Whitney1971}
C.~Whitney, \enquote{{Pauli-Algebraic Operators in Polarization Optics*},} {\protect\JournalTitle{Journal of the Optical Society of America}} \textbf{61}, 1207--1213 (1971).

\bibitem{Wolf2008}
E.~Wolf, \enquote{{Can a light beam be considered to be the sum of a completely polarized and a completely unpolarized beam?}} {\protect\JournalTitle{Optics Letters}} \textbf{33}, 642 (2008).

\bibitem{Tervo2009}
J.~Tervo and J.~Turunen, \enquote{{Comment on “Can a light beam be considered to be the sum of a completely polarized and a completely unpolarized beam?”},} {\protect\JournalTitle{Optics Letters}} \textbf{34}, 1001 (2009).

\bibitem{Piquero2020}
G.~Piquero, R.~Mart{\'{i}}nez-Herrero, J.~C.~G. de~Sande, and M.~Santarsiero, \enquote{{Synthesis and characterization of non-uniformly totally polarized light beams: tutorial},} {\protect\JournalTitle{Journal of the Optical Society of America A}} \textbf{37}, 591 (2020).

\bibitem{Mandel2013}
L.~Mandel and E.~Wolf, \emph{{Optical Coherence and Quantum optics}} (Cambridge University Press, Cambridge, 2013).

\bibitem{Sheppard2016}
C.~J.~R. Sheppard, \enquote{{Parametrization of the Mueller matrix},} {\protect\JournalTitle{Journal of the Optical Society of America A}} \textbf{33}, 2323--2332 (2016).

\end{thebibliography}


\bibliographyfullrefs{reference}

\end{document}